\documentstyle[manuscript,aps,axodraw,epsf]{revtex}
\begin{document}

\title{Non-static Dimensional Reduction of $QED_3$ at Finite Temperature}

\author{Ashok Das}
\address{Department of Physics and Astronomy, University of Rochester,
Rochester, NY 14627}
\author{Gerald Dunne}
\address{Department of Physics, University of Connecticut, Storrs, CT 06269}
%\date{\today}
\maketitle

\begin{abstract}
We study an extreme non-static limit of 2+1-dimensional QED obtained by making a dimensional reduction so that all fields are spatially uniform but time dependent. This dimensional reduction leads to a 0+1-dimensional field theory that inherits many of the features of the 2+1-dimensional model, such as Chern-Simons terms, time-reversal violation, an analogue of parity violation, and global U(2) flavor symmetry. At one-loop level, interactions induce a Chern-Simons term at finite T with coefficient $\tanh(\frac{\beta m_F}{2})$, where $m_F$ is the fermion mass. The finite temperature two loop self-energies are also computed, and are non-zero for all temperatures.
\end{abstract}

\vskip .5in


\pagebreak
\section{Introduction}
The study of induced Chern-Simons terms in $2+1$-dimensional field theory at finite temperature has produced some interesting new insights \cite{dunne1,deser1,schaposnik1,aitchison1,gonzalez,dunne2,barcelos,salcedo,hott} into the relationships between large gauge invariance, effective actions and induced charges at finite temperature. The simplest example of an induced Chern-Simons term arises in the computation of the one-fermion-loop gauge self-energy in $QED_3$. The Chern-Simons coefficient is extracted from the zero-momentum limit of the parity violating part of this self-energy \cite{redlich}. At finite T, this procedure is not unique \cite{kao1} since Feynman diagrams are not analytic in external momenta at finite temperature \cite{weldon}. This is because the thermal heat bath breaks Lorentz invariance, so that diagrams depend separately on the energy $p^0$ and the spatial momentum $\vec{p}$. In a static limit, with $p^0=0$ and $|\vec{p}|\to 0$, an induced Chern-Simons term is found with a temperature dependent coefficient \cite{babu,aitchison2,poppitz}. As first pointed out in \cite{pisarski1}, this result appears (when carried over to a nonabelian theory) to violate large gauge invariance since the coefficient of the induced Chern-Simons term in a nonabelian theory should take discrete values \cite{deser2}. This puzzle has been resolved for the cases when the background has the character of a static magnetic field with integer flux $\Phi$ (as is appropriate for comparison with the static limit), because in this case the problem factorizes into $\Phi$ copies of an exactly solvable $0+1$-dimensional model \cite{dunne1,deser1,schaposnik1}. Then one finds that the finite temperature effective action has an infinite series of parity-violating terms (of which the Chern-Simons term is only the first), each of which has a T dependent coefficient at finite T. Nevertheless, the series is such that the full effective action changes under a large gauge transformation in a manner that is independent of T. These new parity-violating terms are non-extensive (i.e., they are not integrals of a density) and they explicitly vanish at zero temperature (as they must since the zero T effective action should be extensive). This issue is considerably more difficult to resolve in genuinely time-dependent backgrounds, although presumably it should be true that the full finite T effective action is invariant under large gauge transformations that have non-zero winding number. \footnote{Note that the large gauge transformations considered in the static cases have zero winding number; they shift the Chern-Simons action by a constant because of the total derivative term in the change of the Chern-Simons Lagrange density, not because of the winding number density term \cite{lhl}.} 

With this as motivation, we propose here to study an extreme non-static limit of $2+1$-dimensional QED obtained by making a dimensional reduction so that all fields are spatially uniform (i.e. $\vec{p}=0$), but time dependent ($p^0\neq 0$). This dimensional reduction leads to a $0+1$-dimensional field theory (i.e. quantum mechanics) that inherits many of the features of the $2+1$-dimensional model, such as Chern-Simons terms, time-reversal violation, an analogue of parity violation, and global $U(2)$ flavor symmetry \cite{comment}. However, this dimensionally reduced $0+1$-dimensional model is very different from the $0+1$ model considered in \cite{dunne1}. Rather, our dimensionally reduced model is very similar to Chern-Simons quantum mechanics \cite{csqm}, but now coupled to spinor fields. Surprisingly, we find that the interactions induce a Chern-Simons terms at finite T, and from the self-energy diagram we find a coefficient $\tanh(\frac{\beta m_F}{2})$, where $m_F$ is the fermion mass, just as in $2+1$ dimensions. Moreover, the scalar fields are now dynamical and so it is possible to go beyond the one-loop level. However, the analogy with $QED_3$ is not perfect; for example, by suppressing all spatial fluctuation degrees of freedom we have lost local gauge invariance - instead we are left with a type of Yukawa model. Nevertheless, the model displays a remarkably rich renormalization structure, at both zero and non-zero temperature.  

In Section II we introduce the dimensionally reduced model and its symmetry properties. In Section III we calculate the exact one-fermion-loop scalar effective action for the special case where there are no ${\cal T}$ violating bare masses. Section IV introduces the finite temperature propagators, and in Sections V and VI we compute the one-loop scalar self-energy and multi-leg diagrams contributing to the effective action. In Section VII we calculate the one-loop fermion self-energy, and in Section VIII we present some two-loop results. Section IX contains some concluding remarks.

\section{Dimensionally Reduced Model}

Consider the $2+1$ dimensional Lagrange density for quantum electrodynamics together with a Chern-Simons term:
\begin{eqnarray}
{\cal L}_{2+1}(\vec{x},t)=\frac{1}{2}\vec{E}^2-\frac{1}{2}B^2 +\frac{\kappa}{2}\epsilon^{\mu\nu\rho}A_\mu\partial_\nu A_\rho +i\bar{\psi}\gamma^\mu \partial_\mu \psi -m_F\bar{\psi}\psi 
+e\bar{\psi}\gamma^\mu \psi A_\mu
\label{lag3}
\end{eqnarray}
Here, the electric and magnetic fields are $E_i=F_{i0}=\partial_i A_0-\dot{A_i}$, and $B=F_{12} =\epsilon^{ij}\partial_i A_j$, respectively, and $\psi=\left(\matrix{\psi_1\cr\psi_2}\right)$ is a two-component complex spinor field. The Dirac matrices are chosen to be
\begin{eqnarray}
\gamma^0=\left(\matrix{1&0\cr 0&-1}\right),\qquad\qquad \gamma^1=\left(\matrix{0&i\cr i&0}\right), \qquad\qquad
\gamma^2=\left(\matrix{0&1\cr -1&0}\right)
\label{dirac}
\end{eqnarray}
which satisfy $\gamma^\mu \gamma^\nu= \eta^{\mu\nu}-i\epsilon^{\mu\nu\rho}\gamma_\rho$, where $\eta$ is the Minkowski metric $\eta^{\mu\nu}={\rm diag}(1,-1,-1)$. Note that both the fermion mass term $m_F\bar{\psi}\psi$ and the Chern-Simons term $\frac{\kappa}{2}\epsilon^{\mu\nu\rho}A_\mu\partial_\nu A_\rho$ break the discrete symmetries of parity ${\cal P}$ and time reversal ${\cal T}$ in $2+1$ dimensions \cite{deser2,qed3,lhl}. Furthermore, as is well known, a Chern-Simons term for the gauge field can be induced radiatively at zero temperature, as the one-fermion loop gauge self-energy acquires a ${\cal P}$-odd and ${\cal T}$-odd contribution \cite{redlich}. At finite temperature this calculation is more complicated, as was discussed in the introduction.

Now consider a dimensional reduction of the model described by the Lagrange density (\ref{lag3}), defined by taking all fields to be spatially constant. In other words, all momenta $\vec{p}$ are set equal to zero. This reduces to a Lagrangian for a $0+1$ dimensional model :
\begin{eqnarray}
L(t)=\frac{1}{2}\dot{A}_i^2 +\frac{\kappa}{2}\epsilon^{ij}\dot{A}_i A_j +i\bar{\psi} \gamma^0\partial_0\psi  -m_F\bar{\psi}\psi +\mu \psi^\dagger\psi 
+e\bar{\psi} \gamma^i \psi A_i
\label{lag}
\end{eqnarray}
In this spatially constant limit, the non-dynamical field $e A_0$ reduces to a constant chemical potential $\mu$. The vector potential $\vec{A}$ becomes a pair of scalar fields $A_i=A_i(t)$, with a Lorentz-force term $\frac{\kappa}{2}\epsilon^{ij}\dot{A}_i A_j$ inherited from the Chern-Simons mass term, and a Yukawa interaction $e\bar{\psi} \gamma^i \psi A_i$ with the two-component spinor $\psi=\psi(t)$. 

The dimensionally reduced system with Lagrangian (\ref{lag}) retains some interesting symmetry properties. The first is a global rotational invariance, with infinitesimal transformations
\begin{eqnarray}
A_i\to A_i+\lambda \epsilon_{ij}A_j, \qquad\qquad \psi\to(1+i\frac{\lambda}{2}\gamma^0)\psi
\label{rot}
\end{eqnarray}
where $\lambda$ is constant. This clearly leaves each term in the Lagrangian (\ref{lag}) invariant.

The $0+1$-dimensional model in (\ref{lag}) possesses remnants of the behavior of the $QED_3$ Lagrangian (\ref{lag3}) under discrete symmetries \cite{qed3,lhl}. Time reversal acts as
\begin{eqnarray}
{\cal T}:\psi\to \gamma^2\psi,\qquad\qquad {\cal T}:A_i\to -A_i
\label{time}
\end{eqnarray}
This leaves invariant the chemical potential term $\psi^\dagger\psi$, but changes the sign of the mass term $\bar{\psi}\psi$. Similarly, the Chern-Simons term $\epsilon^{ij}\dot{A_i}A_j$ changes sign under ${\cal T}$. The Yukawa interaction term remains invariant (recall that ${\cal T}$ is an antiunitary operation, for which $i\to -i$). All these properties mirror exactly the behavior of the $2+1$ model. 

The $0+1$ model also inherits a charge conjugation symmetry from the $2+1$ model:
\begin{eqnarray}
{\cal C}:\psi\to\psi_c=-\gamma^2\gamma^0\psi^*=
\left(\matrix{\psi_2^*\cr\psi_1^*}\right), \qquad\qquad {\cal C}:A_i\to -A_i
\label{charge}
\end{eqnarray}
Under this generalized charge conjugation operation ${\cal C}$, the chemical potential term $\psi^\dagger\psi$ changes sign, while the mass term $\bar{\psi}\psi$ is invariant (recall that the $\psi$ are anticommuting Fermi fields). Finally, this model also inherits another discrete symmetry, which we still call ``parity'' (although there is no spatial reflection involved), from the parity symmetry of the $2+1$ model:
\begin{eqnarray}
{\cal P}:\psi\to \gamma^1\psi, \qquad {\cal P}:A_1\to -A_1,\qquad 
{\cal P}:A_2\to A_2
\label{parity}
\end{eqnarray}
This symmetry is broken by the fermion mass term $\bar{\psi}\psi$ and the Chern-Simons term $\epsilon^{ij}\dot{A}_i A_j$, but not by the chemical potential term $\psi^\dagger\psi$.

We can also identify a global $U(2)$ flavor symmetry of the model (\ref{lag}), in analogy to the situation in $2+1$ dimensions \cite{qed3,volodya}. The {\it irreducible} Dirac matrix in $0+1$ dimensions is just a number: $\gamma^0=1$. But by reducing from $2+1$ we have effectively doubled the fermionic degrees of freedom (so that we can interpret $\psi_1$ and $\psi_2$ as two flavors) and defined a {\it reducible} Dirac matrix $\gamma^0=\left(\matrix{1&0\cr 0&-1}\right)$. There then exist two other $2\times2$ matrices that anticommute with $\gamma^0$, and taken together the set $\{\gamma^0$, $i\gamma^1$, $i\gamma^2$, ${\bf 1}\}$ generates a global $U(2)$ flavor symmetry. This is exactly analogous to the situation in $2+1$ dimensions where we can double the degrees of freedom taking reducible $4\times 4$ Dirac matrices and two fermion flavors \cite{qed3}. 

This global flavor symmetry acts on the fields as (infinitesimally)
\begin{eqnarray} 
\psi\to (1+\vec{\alpha}\cdot\vec{\gamma})\psi, \qquad\qquad A_i\to A_i+\epsilon_{ij} \alpha^j
\label{flavor}
\end{eqnarray}
This flavor symmetry is broken $U(2)\to U(1)\times U(1)$ by the fermion mass term $\bar{\psi}\psi$, but is preserved by the chemical potential term $\psi^\dagger\psi$. Once again, this mimics the behavior of the $2+1$ dimensional model.

\section{Exact One-loop Effective Action}

In the absence of the bare ${\cal T}$-violating mass terms $\bar{\psi}\psi$ and $\epsilon^{ij}\dot{A}_i A_j$, it is possible to compute the {\it exact} finite temperature one-fermion-loop effective action for the scalar fields $A_i(t)$. Note that the background $A_i(t)$ is manifestly non-static. In this section we consider the bare Lagrangian (obtained from (\ref{lag}) by setting $m_F=0$ and $\kappa=0$)
\begin{eqnarray}
L_0=\frac{1}{2}\dot{A}_i^2 +i\bar{\psi} \gamma^0\partial_0\psi 
+\mu \psi^\dagger\psi +e\bar{\psi} \gamma^i \psi A_i
\label{tlag}
\end{eqnarray}
Integrating out the fermion fields, the one-loop effective action is
\begin{eqnarray}
\Gamma[A_i]=-i \log \left[{\det(i\partial_0+\mu+e\gamma^0\gamma^iA_i)\over \det(i\partial_0+\mu)}\right]
\label{eff}
\end{eqnarray}
We can simplify the evaluation of $\Gamma[A_i]$ by the following nonlinear field redefinition:
\begin{eqnarray}
\psi\to \tilde{\psi}=\exp\left(-ie\int^tdt^\prime \gamma^0\gamma^i A_i\right) \psi
\label{red}
\end{eqnarray}
Then the Lagrangian simplifies to a free Lagrangian $L=\frac{1}{2}\dot{A}_i^2 +\bar{\tilde{\psi}} \gamma^0(i\partial_0+\mu)\tilde{\psi}$. At finite temperature, if we work in the imaginary time formalism, the original fermion fields satisfy antiperiodic boundary conditions: $\psi(\beta)=-\psi(0)$, where $\beta=\frac{1}{T}$ is the inverse temperature. Therefore, the transformed fields $\tilde{\psi}$ have boundary conditions:
\begin{eqnarray}
\tilde{\psi}(\beta)=-\exp\left(-ie\int_0^\beta dt \gamma^0\gamma^i A_i\right) \tilde{\psi}(0)
\label{newperiod}
\end{eqnarray}
This boundary condition may be imposed, for a given background $A_i(t)$, by using a ``chemical potential'' $\tilde{\mu}=i\frac{e}{\beta} \gamma^0\gamma^ia_i$ where
\begin{eqnarray}
a_i=\int_0^\beta dt A_i(t)
\label{smalla}
\end{eqnarray}
Thus, the evaluation of the effective action (\ref{eff}) reduces to the evaluation of the logarithm of the determinant of the operator 
\begin{eqnarray}
i\partial_0+\mu+\tilde{\mu}=\left(\matrix{i\partial_0+\mu& -\frac{e}{\beta}a_-\cr
\frac{e}{\beta}a_+ &i\partial_0+\mu}\right)
\label{op}\end{eqnarray} 
with anti-periodic boundary conditions. Here we have introduced a convenient shorthand notation $a_\pm=a_1\pm ia_2$. It is a straightforward matter to diagonalize this operator by a unitary transformation $U$:
\begin{eqnarray}
i\partial_0+\mu+\tilde{\mu}\to U^\dagger\left(\matrix{i\partial_0+\mu-i\frac{e}{\beta}(a_+a_-)^{1/2} & 0\cr
0&i\partial_0+\mu+i\frac{e}{\beta}(a_+a_-)^{1/2}}\right)U
\label{diagonal}
\end{eqnarray} 
Having the operator in this simple diagonal form, we can now use the results of \cite{dunne1} to evaluate $\Gamma[A_i]$:
\begin{eqnarray}
\Gamma[A_i]&=& -i\left(\log
\left[{\det(i\partial_0+\mu-i\frac{e}{\beta}(a_+a_-)^{1/2})\over \det(i\partial_0+\mu)}\right]
+\log
\left[{\det(i\partial_0+\mu+i\frac{e}{\beta}(a_+a_-)^{1/2})\over \det(i\partial_0+\mu)}\right]\right)\nonumber\\
&=& -i \left(\log\left[\cos(\frac{e}{2}(a_+a_-)^{1/2})-i\tanh(\frac{\beta\mu}{2}) \sin(\frac{e}{2}(a_+a_-)^{1/2})\right] \right.\nonumber\\
&&\qquad \qquad  +\left.\log\left[\cos(\frac{e}{2}(a_+a_-)^{1/2})+i\tanh(\frac{\beta\mu}{2}) \sin(\frac{e}{2}(a_+a_-)^{1/2})\right]\right)\nonumber\\
&=& -i\log\left[\cos^2\left(\frac{e}{2}\left|\int_0^\beta dt A_i(t)\right| \right) +\tanh^2(\frac{\beta\mu}{2}) \sin^2\left(\frac{e}{2}\left| \int_0^\beta dt A_i(t)\right| \right)\right]
\label{exact}
\end{eqnarray}
where $\left| \int_0^\beta dt A_i(t)\right|=\sqrt{a_+ a_-}=\sqrt{a_ia_i}$.

Several comments are in order concerning the result (\ref{exact}). First, 
notice that there is no sign of any ${\cal T}$-violating terms being generated. This is not surprising, since the bare Lagrangian (\ref{tlag}) considered in this section contains neither a fermion mass term nor a Chern-Simons term. Second, this finite temperature effective action involves non-extensive terms, as had also been found in the finite temperature static limit \cite{dunne1,dunne2}. In fact, the result (\ref{exact}) can be expanded as a series in even powers of the integral $\int_0^\beta dt A_i(t)$. At zero temperature, where the effective action must be extensive, the effective action vanishes, by a cancellation of extensive terms $\int_0^\beta dt A_i(t)$ with opposite signs.  

\section{Finite Temperature Perturbation Theory}

While it is interesting to observe the presence of non-extensive terms in the finite temperature effective action for the ${\cal T}$-preserving Lagrangian (\ref{tlag}), it is more interesting to consider the possibility of the radiative generation of ${\cal T}$-violating terms at finite temperature. We expect that this phenomenon may occur if we include the 
${\cal T}$-violating bare masses $m_F$ and/or $\kappa$ that appear in the Lagrangian (\ref{lag}). But the effective action for this system is not exactly solvable. Thus we must resort to finite temperature perturbation theory \cite{das}. Since we are considering non-static backgrounds $A_i=A_i(t)$ it is more direct to use the real-time formalism. We will present the calculations in both coordinate space and in momentum space. 

The finite T propagators simplify considerably in $0+1$ dimensions, due to the simplicity of the dispersion relation. For the fermion fields, the quadratic Lagrangian
\begin{eqnarray}
L_{\rm quad}^{F}=i\bar{\psi} \gamma^0\partial_0\psi  -m_F\bar{\psi}\psi
\label{qf}
\end{eqnarray}
leads to the following thermal real-time propagators (we are using the closed time path formalism \cite{das}):
\begin{eqnarray}
S_{++}(p,m_F)&=&\left(\matrix{
\frac{i}{p-m_F+i\epsilon}-2\pi n_F(m_F)\delta(p-m_F) &0\cr
0& \frac{-i}{p+m_F-i\epsilon}-2\pi n_F(m_F)\delta(p+m_F)}\right)\nonumber\\ \nonumber\\
S_{+-}(p,m_F)&=&\left(\matrix{
-2\pi n_F(m_F)\delta(p-m_F) &0\cr
0& 2\pi (1-n_F(m_F))\delta(p+m_F)}\right)\nonumber\\
\nonumber\\
S_{-+}(p,m_F)&=&\left(\matrix{
2\pi (1-n_F(m_F))\delta(p-m_F) &0\cr
0& -2\pi n_F(m_F)\delta(p+m_F)}\right)\nonumber\\
\nonumber\\
S_{--}(p,m_F)&=&\left(\matrix{
\frac{-i}{p-m_F-i\epsilon}-2\pi n_F(m_F)\delta(p-m_F) &0\cr
0& \frac{i}{p+m_F+i\epsilon}-2\pi n_F(m_F)\delta(p+m_F)}\right)
\label{pp}
\end{eqnarray}
Here, $n_F(m_F)$ is the standard Fermi distribution function
\begin{equation}
n_F(m_F)={1\over e^{\beta m_F}+1}
\label{fermi}
\end{equation} 
In coordinate space, the corresponding propagators $S(t,m)=\int\frac{dp}{2\pi}e^{-ipt}S(p,m)$ are:
\begin{eqnarray}
S_{++}(t,m_F)&=&\left(\matrix{
(\theta(t)-n_F(m_F))e^{-im_F t} &0\cr
0& (\theta(-t)-n_F(m_F))e^{im_F t}}\right)\nonumber\\ \nonumber\\
S_{+-}(t,m_F)&=&\left(\matrix{
-n_F(m_F)e^{-im_F t} &0\cr
0& (1-n_F(m_F))e^{im_F t}}\right)\nonumber\\
\nonumber\\
S_{-+}(t,m_F)&=&\left(\matrix{
(1-n_F(m_F))e^{-im_F t} &0\cr
0& -n_F(m_F)e^{im_F t}}\right)\nonumber\\
\nonumber\\
S_{--}(t,m_F)&=&\left(\matrix{
(\theta(-t)-n_F(m_F))e^{-im_F t} &0\cr
0& (\theta(t)-n_F(m_F))e^{im_F t}}\right)
\label{pt}
\end{eqnarray} 
For the scalar fields, we introduce a mass $m_B$ as an infrared regulator, so that the quadratic Lagrangian
\begin{eqnarray}
L_{\rm quad}^B=\frac{1}{2}\dot{A}_i^2 -\frac{1}{2}m_B^2 A_i^2
\label{qb}
\end{eqnarray}
leads to the following real-time thermal propagators $G^{ij}_{ab}(p,m_B)=\delta_{ij}G_{ab}(p,m_B)$, where $a,b$ refer to the thermal indices $\pm$:
\begin{eqnarray}
G_{++}(p,m_B)&=&{i\over p^2-m_B^2+i\epsilon}+2\pi n_B(m_B) \delta(p^2-m_B^2)\nonumber\\
G_{+-}(p,m_B)&=&2\pi [\theta(-p)+n_B(m_B)] \delta(p^2-m_B^2)\nonumber\\
G_{-+}(p,m_B)&=&2\pi [\theta(p)+n_B(m_B)]  \delta(p^2-m_B^2)\nonumber\\
G_{--}(p,m_B)&=&{-i\over p^2-m_B^2-i\epsilon}+2\pi n_B(m_B) \delta(p^2-m_B^2)
\label{bpp}
\end{eqnarray}
Here, $n_B(m_B)$ is the standard Bose distribution function
\begin{equation}
n_B(m_B)={1\over e^{\beta m_B}-1}
\label{bose}
\end{equation} 
In coordinate space the corresponding propagators are
\begin{eqnarray}
G_{++}(t,m_B)&=&\frac{1}{2m_B}\left( [\theta(t)+n_B(m_B)]e^{-im_B t}+ [\theta(-t)+n_B(m_B)]e^{im_B t}\right) \nonumber\\
G_{+-}(t,m_B)&=&\frac{1}{2m_B}\left( n_B(m_B)e^{-im_B t}+ [1+n_B(m_B)] 
e^{im_B t}\right) \nonumber\\
G_{-+}(t,m_B)&=&\frac{1}{2m_B}\left( [1+n_B(m_B)]e^{-im_B t}+ n_B(m_B)
e^{im_B t}\right) \nonumber\\
G_{--}(t,m_B)&=&\frac{1}{2m_B}\left( [\theta(-t)+n_B(m_B)]e^{-im_B t}+ [\theta(t)+n_B(m_B)]e^{im_B t}\right)
\label{bpt}
\end{eqnarray}

\section{One-fermion-loop Scalar Self-energy}

Given the finite T propagators from the previous section, we now compute the one-fermion-loop self-energy for the scalar fields $A_i$. We present the calculation both in momentum space and in coordinate space.

\subsection{Momentum Space Calculation}

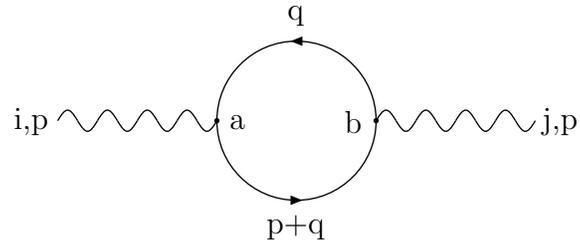
\begin{figure}
\centering{
\begin{picture}(200,200)(0,0)
\Photon(10,100)(70,100)4 4
\Photon(130,100)(190,100)4 4
\ArrowArc(100,100)(30,0,180)
\ArrowArc(100,100)(30,180,360)
\Vertex(70,100)1
\Vertex(130,100)1
\Text(0,100)[c]{i,p}
\Text(200,100)[c]{j,p}
\Text(100,140)[c]{q}
\Text(100,60)[c]{p+q}
\Text(75,100)[l]{a}
\Text(125,100)[r]{b}
\end{picture}}
\caption{The scalar self energy $\Pi_{ab}^{ij}(p)$ in momentum space. The external momentum is $p$, the external field indices are $i$ and $j$, and the thermal indices are $a$ and $b$.}
\label{f1}
\end{figure}

The momentum space scalar self-energy with thermal indices $a,b$ (each taking possible values $\pm$) in Figure \ref{f1} is (this definition includes a negative sign inherent in the definition of $\Pi_{\pm\mp}$ in the closed time path formalism \cite{das})
\begin{eqnarray}
\Pi^{ij}_{ab}(p)= -(-ie)^2\int{dq\over 2\pi} {\rm Tr} \left(\gamma^i S_{ab}(q,m_F) \gamma^j S_{ba}(p+q,m_F)\right)
\label{sse}
\end{eqnarray}
The Dirac trace may be simplified with the useful algebraic identity
\begin{equation}
\gamma^i \left(\matrix{c&0\cr 0&d}\right) \gamma^j =-\left(\matrix{(\delta^{ij}+i\epsilon^{ij}) d &0\cr 0& (\delta^{ij}-i\epsilon^{ij}) c}\right)
\label{id}
\end{equation}
Thus we see that the self-energy naturally separates into a piece proportional to $\delta^{ij}$ and a piece proportional to $\epsilon^{ij}$. We first consider the $++$ thermal self-energy:
\begin{eqnarray}
\Pi^{ij}_{++}(p)&=&-e^2\left\{ \delta^{ij} \int{dq\over 2\pi} S^{11}_{++}(q,m_F)\left[S_{++}^{22}(q+p,m_F)+S_{++}^{22}(q-p,m_F)\right]\right.
\nonumber\\
&&\qquad\qquad \left. -i\epsilon^{ij} \int{dq\over 2\pi} S^{11}_{++}(q,m_F)\left[S_{++}^{22}(q+p,m_F)-S_{++}^{22}(q-p,m_F)\right]\right\}\label{se}
\end{eqnarray}
Here the upper indices $11$ and $22$ refer to the diagonal components of the Dirac matrix structure of the fermionic propagators in (\ref{pp}), and we have used the fact that $S_{++}^{22}(-q,m_F)=S_{++}^{11}(q,m_F)$. Notice that the part of $\Pi^{ij}_{++}(p)$ that is proportional to $\delta^{ij}$ is manifestly even in $p$, while the part proportional to $\epsilon^{ij}$ is odd in $p$. The loop momentum integrals are straightforward contour integrals, and we find the remarkably simple structure for the self-energy
\begin{eqnarray}
\Pi^{ij}_{++}(p)&=&-e^2 \tanh(\frac{\beta m_F}{2})\left[4 m_F\delta^{ij} -2ip\epsilon^{ij}\right] \left({i\over p^2-4m_F^2+i\epsilon}+2\pi n_B(2m_F)\delta(p^2-4m_F^2)\right)\nonumber\\
&=& -e^2 \tanh(\frac{\beta m_F}{2})\left[4m_F\delta^{ij} -2ip\epsilon^{ij}\right] G_{++}(p,2m_F)
\label{result}
\end{eqnarray}
To obtain this result we have used the identities:
\begin{eqnarray}
\delta(p-2 m_F)+\delta(p+2 m_F)&=& 4 m_F\delta(p^2-4 m_F^2)\nonumber\\
\delta(p-2 m_F)-\delta(p+2 m_F)&=& 2p\delta(p^2-4 m_F^2) 
\label{deltas}
\end{eqnarray}
Furthermore, the $\tanh(\frac{\beta m_F}{2})$ factor arises because of the identity
\begin{equation}
[n_F(m_F)]^2=\tanh(\frac{\beta m_F}{2})\, n_B(2 m_F)
\label{m1}
\end{equation}
The result (\ref{result}) is interesting for a number of reasons. First, notice the appearance of a Chern-Simons-like $\epsilon^{ij}$ part to the scalar self-energy. Furthermore, notice that the overall factor 
$\tanh(\frac{\beta m_F}{2})$ that is familiar from the computation of induced Chern-Simons terms at finite temperature in $2+1$ dimensional theories in the static limit \cite{babu,aitchison2}. Indeed, in the small $p$ limit, 
\begin{eqnarray}
\Pi^{ij}_{++}(p\to 0)\sim  \tanh(\frac{\beta m_F}{2})\left[i\frac{e^2}{m_F} \delta^{ij} +\frac{e^2}{2 m_F^2} p\epsilon^{ij}\right] 
\label{limit}
\end{eqnarray}
which shows that at one-loop order we induce both an even and odd mass term for the scalars $A_i$. 

It is also interesting to note that the full $p$ dependence of the self-energy combines into a bosonic propagator, for a bosonic field of mass $2m_F$, even though the actual calculation of the fermion loop only involves fermionic propagators. We also comment that if this calculation is repeated including also a chemical potential term $\mu \psi^\dagger \psi$ for the fermions, then the only difference in the final result is that
\begin{equation}
\tanh(\frac{\beta m_F}{2})\qquad {\rm becomes} \qquad {\sinh(\beta m_F)\over \cosh(\beta m_F)+\cosh(\beta \mu)}
\label{cp}
\end{equation}
which arises because of the identity
\begin{equation}
n_F(m_F+\mu) n_F(m_F-\mu)= \left({\sinh(\beta m_F)\over 
\cosh(\beta m_F)+\cosh(\beta \mu)}\right) n_B(2 m_F)
\label{rel}
\end{equation}
This modified prefactor is precisely the Chern-Simons prefactor found in the static limit finite temperature calculation in $2+1$ dimensions with both a fermion mass $m_F$ and a chemical potential $\mu$ \cite{poppitz}. 

This momentum space calculation can be done similarly for the other thermal components of the self-energy, with the general result that
\begin{eqnarray}
\Pi^{ij}_{ab}(p)= -e^2 \tanh(\frac{\beta m_F}{2})\left[4 m_F\delta^{ij} -2ip\epsilon^{ij}\right] G_{ab}(p,2m_F)
\label{gen}
\end{eqnarray}
where, as stated before, this definition includes a negative sign inherent in the definition of $\Pi_{\pm\mp}$ in the closed time path formalism \cite{das}.

\subsection{Coordinate Space Calculation}

The one-fermion-loop scalar self-energy is even easier to compute in coordinate space (see Figure \ref{f2}) as no integration is involved - it is simply a matter of multiplying out the various coordinate space propagators. Thus, for the $++$ thermal component one finds 
\begin{eqnarray}
\Pi^{ij}_{++}(t_1-t_2)&=& -(ie)^2 {\rm Tr} \left(\gamma^i S_{++}(t_1-t_2,m_F) \gamma^j S_{++}(t_2-t_1,m_F)\right)\nonumber\\
&&\hskip -1in =-e^2\left\{ \delta^{ij} \left[ (\theta(t_2-t_1)-n_F(m_F))^2 
e^{2i m_F (t_1-t_2)} +(\theta(t_1-t_2)-n_F(m_F))^2 
e^{-2i m_F (t_1-t_2)}\right]\right.\nonumber\\
&&\hskip -1in \qquad \left. +i\epsilon^{ij} \left[ (\theta(t_2-t_1)-n_F(m_F))^2 
e^{2i m_F (t_1-t_2)} -(\theta(t_1-t_2)-n_F(m_F))^2 
e^{-2i m_F (t_1-t_2)}\right]\right\}
\label{ssc}
\end{eqnarray}
As before, we recognize a term proportional to $\delta^{ij}$ and a term proportional to $\epsilon^{ij}$. The former is even under the interchange of $t_1$ and $t_2$, while the latter is odd under this interchange. The products of propagators may be expanded using the identity
\begin{equation}
(\theta(t)-n_F(m_F))^2=\tanh(\frac{\beta m_F}{2})\, (\theta(t)+n_B(2m_F))
\label{cid}
\end{equation}
Thus, we see immediately that
\begin{eqnarray}
\Pi^{ij}_{++}(t_1-t_2)= -e^2 \tanh(\frac{\beta m_F}{2})\left[ 4 m_F\delta^{ij}+2\epsilon^{ij}\frac{\partial}{\partial t_1} \right] G_{++}(t_1-t_2,2m_F)
\label{cresult}
\end{eqnarray}
which is simply the coordinate space version of the result (\ref{result}). The other thermal components are computed similarly, so that 
\begin{eqnarray}
\Pi^{ij}_{ab}(t_1-t_2)= -e^2 \tanh(\frac{\beta m_F}{2})\left[ 4 m_F\delta^{ij}+2\epsilon^{ij}\frac{\partial}{\partial t_1} \right] G_{ab}(t_1-t_2,2m_F)
\label{gencresult}
\end{eqnarray}
in agreement with the momentum space result (\ref{gen}).

\section{One-loop Effective Action}

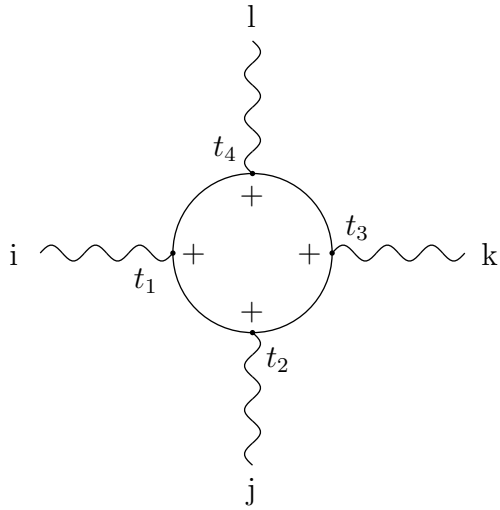
\begin{figure}
\centering{
\begin{picture}(200,200)(0,0)
\Photon(20,100)(70,100)3 3
\Photon(130,100)(180,100)3 3
\Photon(100,130)(100,180)3 3
\Photon(100,70)(100,20)3 3
\CArc(100,100)(30,0,359)
\Vertex(70,100)1
\Vertex(130,100)1
\Vertex(100,130)1
\Vertex(100,70)1
\Text(78,100)[c]{+}
\Text(122,100)[c]{+}
\Text(100,78)[c]{+}
\Text(100,122)[c]{+}
\Text(60,90)[c]{$t_1$}
\Text(110,60)[c]{$t_2$}
\Text(140,110)[c]{$t_3$}
\Text(90,140)[c]{$t_4$}
\Text(10,100)[c]{i}
\Text(100,10)[c]{j}
\Text(190,100)[c]{k}
\Text(100,190)[c]{l}
\end{picture}}
\caption{The four-point function in coordinate space}
\label{f4}
\end{figure}

For the model with a fermion mass term $m_F\bar{\psi}\psi$, it is not possible to compute the one-loop effective action in closed form. Nevertheless, a perturbative analysis is possible. The effective action has the standard diagrammatic expansion shown in Figure \ref{f3}. 
It is clear that the tadpole term vanishes because of the trace property 
${\rm Tr}(\gamma^i S(q,m_F))=0$, since $\gamma^i$ is off-diagonal and $S$ is diagonal. For this same reason, each diagram with an odd number of external scalar lines vanishes identically. (This is analogous to Furry's theorem). Beyond the self-energy diagram, which was computed in the previous section, the next is the four-leg diagram. As noted for the self-energy, it is easiest to do these one-loop calculations in coordinate space since there are no integrations involved. For example, the 4-leg diagram in Figure \ref{f4} yields
\begin{eqnarray}
\Gamma^{ijkl}_{++++}(t_1,t_2,t_3,t_4) &=& -(ie)^4 {\rm Tr}\left( \gamma^i S_{++}(t_1-t_4) \gamma^l S_{++}(t_4-t_3) \gamma^k S_{++}(t_3-t_2) \gamma^j S_{++}(t_2-t_1)\right) \nonumber\\
&&\hskip -1in =-e^4\left[(\delta^{il}+i\epsilon^{il}) (\delta^{kj}+i\epsilon^{kj}) 
S_{++}^{11}(t_4-t_1) S_{++}^{11}(t_4-t_3) S_{++}^{11}(t_2-t_3) S_{++}^{11}(t_2-t_1) \right.\nonumber\\
&&\hskip -1in +\left.
(\delta^{il}-i\epsilon^{il}) (\delta^{kj}-i\epsilon^{kj}) 
S_{++}^{11}(t_1-t_4) S_{++}^{11}(t_3-t_4) S_{++}^{11}(t_3-t_2) S_{++}^{11}(t_1-t_2)\right]
\end{eqnarray}
Adding the other permutations of the external coordinates, the quartic contribution to the effective action can be written as $S_{\rm quartic}=\int dt_1 dt_2 dt_3 dt_4 G^{(4)}(t_1,t_2,t_3,t_4)$ where
\begin{eqnarray}
G^{(4)}&=& {ie^4\over 2 (4!)}\left[ \bar{A}(t_1) A(t_2) \bar{A}(t_3) A(t_4) S_{++}^{11}(t_4-t_1) S_{++}^{11}(t_4-t_3) S_{++}^{11}(t_2-t_3) S_{++}^{11}(t_2-t_1)\right.\nonumber\\
&&+\left.\bar{A}(t_1) A(t_3) \bar{A}(t_4) A(t_2) S_{++}^{11}(t_2-t_1) S_{++}^{11}(t_2-t_4) S_{++}^{11}(t_3-t_1) S_{++}^{11}(t_3-t_4)\right.\nonumber\\
&&+\left.\bar{A}(t_1) A(t_4) \bar{A}(t_2) A(t_3) S_{++}^{11}(t_3-t_1) S_{++}^{11}(t_3-t_2) S_{++}^{11}(t_4-t_1) S_{++}^{11}(t_4-t_2)\right.\nonumber\\
&&+\left. A(t_1)\bar{A}(t_2) A(t_3)\bar{A}(t_4)  S_{++}^{11}(t_1-t_4) S_{++}^{11}(t_1-t_2) S_{++}^{11}(t_3-t_4) S_{++}^{11}(t_3-t_2)\right.\nonumber\\
&&+\left.A(t_1)\bar{A}(t_3) A(t_4)\bar{A}(t_2) S_{++}^{11}(t_1-t_2) S_{++}^{11}(t_1-t_3) S_{++}^{11}(t_4-t_2) S_{++}^{11}(t_4-t_3)\right.\nonumber\\
&&+\left. A(t_1)\bar{A}(t_4) A(t_2)\bar{A}(t_3) S_{++}^{11}(t_1-t_3) S_{++}^{11}(t_1-t_4) S_{++}^{11}(t_2-t_3) S_{++}^{11}(t_2-t_4)\right]
\label{fourth}
\end{eqnarray}
where we have defined $A=A_1+iA_2$ and $\bar{A}=A_1-iA_2$.

The structure is similar for the $2N$-point function, for which we find
\begin{eqnarray}
G^{(2N)}(t_1,\dots,t_{2N})&=& {ie^{2N}\over 2^{N-1} (2N)!}\left[ 
\prod_{i=1}^N \left(\bar{A}(t_{2i-1}) A(t_{2i}) S_{++}^{11}(t_{2i}-t_{2i-1}) S_{++}^{11}(t_{2i}-t_{2i+1})\right) \right.\nonumber\\
&&+\left.\prod_{i=1}^N \left(A(t_{2i-1}) \bar{A}(t_{2i})  S_{++}^{11}(t_{2i-1}-t_{2i}) S_{++}^{11}(t_{2i+1}-t_{2i})\right) \right.\nonumber\\
&&+\left. {\rm cyclic ~ in ~} t_2, t_3, \dots t_{2N}\right]
\label{2nleg}
\end{eqnarray}
Here we have identified $t_{2N+1}\equiv t_1$. 

Notice that this contribution to the effective action involves non-extensive terms, although the zero T limit is extensive as the $2N$-point function reduces to products of theta functions. Also, it is interesting to note that this finite T effective action is considerably more complicated than the situation with a static background, for which it is possible to find the exact ${\cal P}$-odd part of the effective action \cite{dunne1,deser1,schaposnik1}. Here, in a non-static limit, it does not appear possible to find such an exact closed-form expression.

\section{One-loop Fermion Self-energy}

\subsection{Momentum Space Calculation}

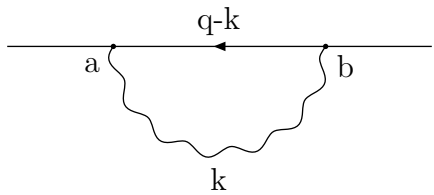
\begin{figure}
\centering{
\begin{picture}(200,100)(0,0)
\ArrowLine(180,80)(20,80)
\PhotonArc(100,80)(40,180,360)2 7
\Vertex(140,80)1
\Vertex(60,80)1
\Text(100,90)[c]{q-k}
\Text(100,30)[c]{k}
\Text(148,73)[c]{b}
\Text(52,73)[c]{a}
\end{picture}}
\caption{The fermion self energy $\Sigma_{ab}(q)$ in momentum space. The external momentum is $q$, and the thermal indices are $a$ and $b$.}
\label{f5}
\end{figure}

The one-loop fermion self-energy in Figure \ref{f5} takes an interesting simple form at finite temperature. In momentum space it is straightforward to perform the single loop integral to show that 
\begin{eqnarray}
\Sigma_{ab}(q)&=&(ie)^2\int \frac{dk}{2\pi} \, \gamma^i S_{ab}(q-k,m_F)\gamma^j G^{ij}_{ab}(k,m_B)\nonumber\\
&=&-\frac{e^2}{m_B}\left\{ [1+n_B(m_B)-n_F(m_F)] \left(\matrix{ S_{ab}^{22}(q,m_F+m_B) &0\cr 0& S_{ab}^{11}(q,m_F+m_B)}\right) \right.\nonumber\\
&&\qquad  +\left. [n_B(m_B)+n_F(m_F)] \left(\matrix{ S_{ab}^{22}(q,m_F-m_B) &0\cr 0& S_{ab}^{11}(q,m_F-m_B)}\right)\right\}
\label{fse}
\end{eqnarray}
Here we have used the identities (we assume the scalar regulator mass $m_B$ is smaller than the fermionic mass $m_F$)
\begin{eqnarray}
{n_F(m_F) n_B(m_B)\over n_F(m_F+m_B)} &=&  [1+n_B(m_B)-n_F(m_F)]\nonumber\\
&=& \frac{1}{2}\left(\coth(\frac{\beta m_B}{2})+\tanh(\frac{\beta m_F}{2})\right)\nonumber\\
{n_F(m_F)[1+ n_B(m_B)]\over n_F(m_F-m_B)} &=& [n_B(m_B)+n_F(m_F)]\nonumber\\
&=& \frac{1}{2}\left(\coth(\frac{\beta m_B}{2})-
\tanh(\frac{\beta m_F}{2})\right)
\label{fids}
\end{eqnarray}
It is interesting to notice that the result (\ref{fse}) says that the effect of the scalar loop is essentially to split the bare fermion (of mass $m_F$) into two fermions, of masses $m_F\pm m_B$, with weighting factors $\frac{1}{2}\left(\coth(\frac{\beta m_B}{2})\pm \tanh(\frac{\beta m_F}{2})\right)$ respectively. This, in turn, makes higher-loop calculations easier, since a self-energy insertion is simply like an insertion of a modified propagator.

\subsection{Coordinate Space Calculation} 

The same result is found in coordinate space. For example, the $++$ thermal component of the fermion self-energy is simply found by multiplying out the following propagators
\begin{eqnarray}
\Sigma_{++}(t_1-t_2)&=&(ie)^2\, \gamma^i S_{++}(t_1-t_2,m_F)\gamma^j G^{ij}_{++}(t_1-t_2,m_B)\nonumber\\
&&\hskip -1in =\frac{e^2}{m_B}\left(
\matrix{ [\theta(t_2-t_1)-n_F(m_F)]e^{im_F(t_1-t_2)}  &0\cr 0&  [\theta(t_1-t_2)-n_F(m_F)]e^{-im_F(t_1-t_2)}}\right) \nonumber\\
&&\hskip -1in \qquad\qquad  \times 
\left\{ [\theta(t_1-t_2)+n_B(m_B)]e^{-im_B(t_1-t_2)}  + [\theta(t_2-t_1)+n_B(m_B)]e^{im_B(t_1-t_2)}\right\}\nonumber\\
&&\hskip -1in =\frac{-e^2}{m_B}\left\{ [1+n_B(m_B)-n_F(m_F)] \left(
\matrix{ S_{++}^{22}(t_1-t_2,m_F+m_B) &0\cr 0& S_{++}^{11}(t_1-t_2,m_F+m_B)} \right) \right.\nonumber\\
&&\hskip -3cm \qquad  +\left. [n_B(m_B)+n_F(m_F)] \left(\matrix{ S_{++}^{22}
(t_1-t_2,m_F-m_B) &0\cr 0& S_{++}^{11}(t_1-t_2,m_F-m_B)}\right)\right\}
\label{fsec}
\end{eqnarray}
Here we have used the identities
\begin{eqnarray}
(\theta (t)-n_F(m_F)) (\theta (t)+n_B(m_B)) &=& (1+n_B(m_B)-n_F(m_F)) 
(\theta (t)-n_F(m_F+m_B)) \nonumber\\
\hskip -5pt(\theta(t) - n_F(m_F)) (\theta (-t)+n_B(m_B)) \hskip -2pt &=&(n_B(m_B)+n_F(m_F)) (\theta (t) - n_F(m_F-m_B))
\label{cidd}
\end{eqnarray}
The analogous results hold for the other thermal indices of the fermion self-energy.

\section{Two-loop results} 

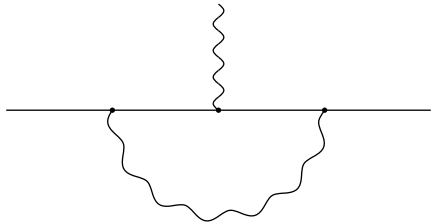
\begin{figure}
\centering{
\begin{picture}(200,120)(0,0)
\Line(180,80)(20,80)
\PhotonArc(100,80)(40,180,360)2 7
\Photon(100,80)(100,120)2 4
\Vertex(140,80)1
\Vertex(60,80)1
\Vertex(100,80)1
\end{picture}}
\caption{The one-loop vertex function. This vertex correction vanishes due to Dirac matrix structure.}
\label{f7}
\end{figure}

\begin{figure}
\centering{
\begin{picture}(400,300)(0,0)
\Photon(10,200)(70,200)4 4
\Photon(130,200)(190,200)4 4
\PhotonArc(100,242)(30,225,315)4 4
\CArc(100,200)(30,0,180)
\CArc(100,200)(30,180,360)
\Vertex(70,200)1
\Vertex(130,200)1
\Vertex(79,221)1
\Vertex(121,221)1
\Text(75,200)[l]{+}
\Text(65,190)[c]{$t_1$}
\Text(10,190)[c]{$i$}
\Text(135,190)[l]{$t_2$}
\Text(190,190)[c]{$j$}
\Text(125,200)[r]{+}
\Text(135,225)[c]{$+$,$t_4$}
\Text(65,225)[c]{$+$,$t_3$}
\Text(100,160)[c]{(a)}
\Photon(210,200)(270,200)4 4
\Photon(330,200)(390,200)4 4
\PhotonArc(300,242)(30,225,315)4 4
\CArc(300,200)(30,0,180)
\CArc(300,200)(30,180,360)
\Vertex(270,200)1
\Vertex(330,200)1
\Vertex(279,221)1
\Vertex(321,221)1
\Text(275,200)[l]{+}
\Text(265,190)[c]{$t_1$}
\Text(335,190)[l]{$t_2$}
\Text(325,200)[r]{+}
\Text(335,225)[c]{$-$,$t_4$}
\Text(265,225)[c]{$+$,$t_3$}
\Text(210,190)[c]{$i$}
\Text(390,190)[c]{$j$}
\Text(300,160)[c]{(b)}
\Photon(10,100)(70,100)4 4
\Photon(130,100)(190,100)4 4
\PhotonArc(100,142)(30,225,315)4 4
\CArc(100,100)(30,0,180)
\CArc(100,100)(30,180,360)
\Vertex(70,100)1
\Vertex(130,100)1
\Vertex(79,121)1
\Vertex(121,121)1
\Text(75,100)[l]{+}
\Text(65,90)[c]{$t_1$}
\Text(135,90)[l]{$t_2$}
\Text(125,100)[r]{+}
\Text(135,125)[c]{$+$,$t_4$}
\Text(65,125)[c]{$-$,$t_3$}
\Text(10,90)[c]{$i$}
\Text(190,90)[c]{$j$}
\Text(100,60)[c]{(c)}
\Photon(210,100)(270,100)4 4
\Photon(330,100)(390,100)4 4
\PhotonArc(300,142)(30,225,315)4 4
\CArc(300,100)(30,0,180)
\CArc(300,100)(30,180,360)
\Vertex(270,100)1
\Vertex(330,100)1
\Vertex(279,121)1
\Vertex(321,121)1
\Text(275,100)[l]{+}
\Text(265,90)[c]{$t_1$}
\Text(335,90)[l]{$t_2$}
\Text(325,100)[r]{+}
\Text(335,125)[c]{$-$,$t_4$}
\Text(265,125)[c]{$-$,$t_3$}
\Text(210,90)[c]{$i$}
\Text(390,90)[c]{$j$}
\Text(300,60)[c]{(d)}
\end{picture}}
\caption{The two-loop contributions to the scalar self energy $\Pi_{++}^{ij}(t_1-t_2)$ in coordinate space. Note that the diagrams (b) and (c) have a single -- vertex and so contribute with the opposite sign compared to diagrams (a) and (d).}
\label{f9}
\end{figure}
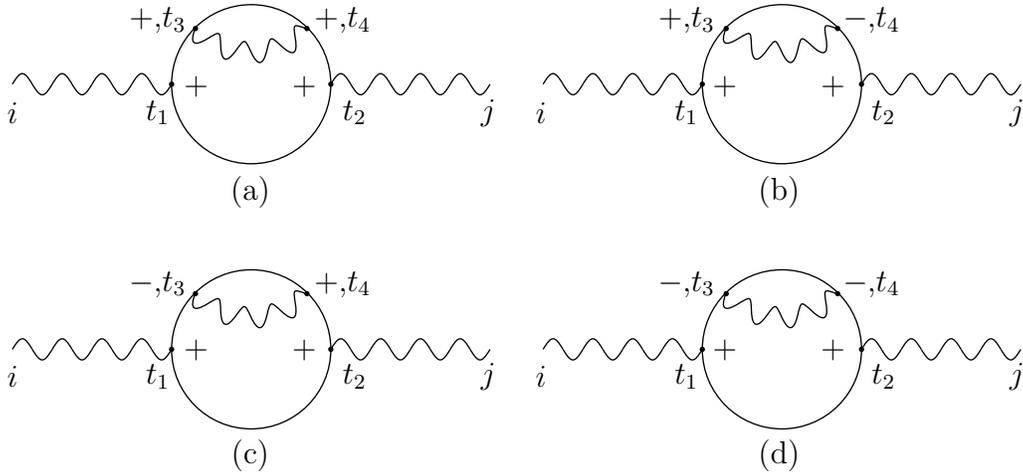

The simple structure of the finite temperature one-loop scalar and fermion self-energies found in the preceding sections suggests that higher loop calculations should be relatively straightforward also. As an example of the continued simplicity, consider the one-loop vertex correction in Figure \ref{f7}. This one-loop vertex correction vanishes identically due to Dirac matrix algebra since it involves the matrix structure
\begin{eqnarray}
\gamma^i \left(\matrix{0& c\cr d&0}\right) \gamma^i =0
\label{vanish}
\end{eqnarray}
This has the immediate consequence that the crossed two-loop scalar self-energy diagram in Figure \ref{f8}(b) vanishes, and so we only need to consider the diagram in Figure \ref{f8}(a). To compute the contribution for the external $++$ thermal indices, we need four such diagrams, as shown in Figure \ref{f9}, since the internal vertices can each have one of two thermal indices $a=\pm$. So
\begin{eqnarray}
{\Pi^{ij}}_{++}(t_1-t_2) &=&-(ie)^4\sum_{a,b}\, \int dt_3 \int dt_4\,
\nonumber \\
&&\hskip-2cm  tr \left[ \gamma^i S_{+a} (t_1-t_3)\gamma^k S_{ab} (t_3-t_4) \gamma^l S_{b+} (t_4-t_2) \gamma^j S_{++} (t_2-t_1)\right] G^{ab}_{kl}(t_3-t_4)
\label{2l}
\end{eqnarray}
But we have already computed part of this, in evaluating the one-loop fermion self-energy $\Sigma$ in the previous section. From the result (\ref{fsec}) we see that $\Sigma$ splits into a contribution proportional to $(1+n_B(m_B)-n_F(m_F))$ and a contribution proportional to $(n_B(m_B)+n_F(m_F))$. Therefore, the combination $S\Sigma S$, which appears in the two-loop calculation, also splits in this manner. These two contributions are simply related by taking $(m_F+m_B)$ to $(m_F-m_B)$. Furthermore, the product $S\Sigma S$ is diagonal in the Dirac matrix sense, and the two diagonal entries are simply related by interchanging the coordinate labels. So it is sufficient to look first at the piece of the $(11)$ Dirac matrix component that is multiplied by $(1+n_B(m_B)-n_F(m_F))$:
\begin{eqnarray}
-(ie)^2 \sum_{a,b}\left(S_{+a} (t_1-t_3)\Sigma_{ab} (t_3-t_4) S_{b+} (t_4-t_2) \right)^{(11)}&=& \nonumber\\
&&\hskip -3in \frac{e^4}{m_B} (1+n_B(m_B)-n_F(m_F)) e^{-im_F(t_1-t_3)+i(m_F+m_B)(t_3-t_4)-im_F(t_4-t_2)}\nonumber\\
&& \hskip -3in\left[ \left(\theta(t_1-t_3)-n_F(m_F)\right) \left\{ \left(\theta(t_4-t_3)-n_F(m_F+m_B)\right)\left(\theta(t_4-t_2)-n_F(m_F)\right) \right.\right.\nonumber\\
&& \hskip -3cm -\left.\left. \left(1-n_F(m_F+m_B)\right)\left(1-n_F(m_F)\right) \right\}\right.\nonumber\\
&& \hskip -3in \left. +n_F(m_F)\left\{ \left(-n_F(m_F+m_B)\right)\left(\theta(t_4-t_2)-n_F(m_F)\right)\right. \right.\nonumber\\
&&\hskip -3cm - \left.\left.\left(\theta(t_3-t_4)-n_F(m_F+m_B)\right)\left(1-n_F(m_F)\right)\right\}\right]
\label{tlmess}
\end{eqnarray}
After some straightforward algebra, this reduces to
\begin{eqnarray}
&&\frac{e^4}{m_B} (1+n_B(m_B)-n_F(m_F))e^{-im_F(t_1-t_2)} e^{i(m_B+2m_F)(t_3-t_4)} \nonumber\\
&&\cdot \left[ \theta(t_2-t_4)\theta(t_4-t_3)\theta(t_3-t_1)+n_F(m_F+m_B) \theta(t_1-t_3)\theta(t_2-t_4)\right.\nonumber\\
&&\left.-(1-n_F(m_F))\left(\theta(t_1-t_3)\theta(t_3-t_4)+ \theta(t_2-t_4)\theta(t_4-t_3)\right)\right]
\label{three}
\end{eqnarray}
Inserting this into the two-loop self-energy (\ref{2l}) requires performing the integrals over the internal vertex coordinates $t_3$ and $t_4$. As can be seen from (\ref{three}), there are three types of integrals, and they require some care since they involve products of theta functions. Using the representation
\begin{eqnarray}
\theta(t)=\int \frac{dk}{-2\pi i}\, {e^{-ikt}\over k+i\epsilon}
\label{reg}
\end{eqnarray}
together with careful regularization of delta functions, we find 
\begin{eqnarray}
\int dt_3 \int dt_4 e^{i(m_B+2m_F) (t_3-t_4)} \theta(t_2-t_4)\theta(t_4-t_3)\theta(t_3-t_1)&&\nonumber\\
&&\hskip -2in =\theta(t_2-t_1)\left[{1-e^{i(m_B+2m_F)(t_1-t_2)}\over (m_B+2m_F)^2}+{i(t_1-t_2)\over (m_B+2m_F)}\right]
\label{int1}
\end{eqnarray}
\begin{eqnarray}
\int dt_3 \int dt_4 e^{i(m_B+2m_F) (t_3-t_4)} \left(\theta(t_1-t_3)\theta(t_3-t_4)+ \theta(t_2-t_4)\theta(t_4-t_3)\right) &&\nonumber\\
&&\hskip -3in ={2\over (m_B+2m_F)^2}+{i(t_1-t_2)\over (m_B+2m_F)}
\label{int2}
\end{eqnarray}
\begin{eqnarray}
\int dt_3 \int dt_4 e^{i(m_B+2m_F) (t_3-t_4)}\theta(t_1-t_3)\theta(t_2-t_4) =\frac{1}{(m_B+2m_F)^2}+{e^{i(m_B+2m_F)(t_1-t_2)}\over (m_B+2m_F)^2}
\label{int3}
\end{eqnarray}
This completes the calculation of all the parts contributing to the $(1+n_B(m_B)-n_F(m_F))$ part of 
\begin{eqnarray}
-(ie)^2 \sum_{a,b}\int dt_3 \int dt_4 \left(S_{+a} (t_1-t_3)\Sigma_{ab} (t_3-t_4) S_{b+} (t_4-t_2) \right)^{(11)}
\end{eqnarray}
The analogous term proprtional to $(n_B(m_B)+n_F(m_F))$ is obtained by replacing $m_F+m_B$ with $m_F-m_B$. And then the $(22)$ Dirac matrix components are obtained by interchanging $t_1$ and $t_2$. 

Finally, to evaluate the two-loop scalar self-energy (\ref{2l}), we insert the remaining gamma matrices and the propagator $S_{++} (t_2-t_1)$, and perform the Dirac trace. The net result is 
\begin{eqnarray}
{\Pi^{ij}}_{++}(t_1-t_2) &=&\frac{e^4}{m_B}\left[ \right.\nonumber\\
&&\hskip -1in (1-2n_F(m_F))\left( {(1+n_B(m_B)-n_F(m_F))\over (m_B+2m_F)^2}+ {(n_B(m_B)+n_F(m_F))\over (m_B-2m_F)^2}\right)\cdot\nonumber\\
&&\hskip 2cm \cdot (\delta^{ij}+\frac{1}{2m_F}\epsilon^{ij}\frac{\partial}{\partial t_1})G_{++}(t_1-t_2,2m_F)\nonumber\\
&&\hskip -1in -(1-2n_F(m_F))\left( {(1+n_B(m_B)-n_F(m_F))\over (m_B+2m_F)}- {(n_B(m_B)+n_F(m_F))\over (m_B-2m_F)}\right)\cdot\nonumber\\
&&\hskip 2cm \cdot (t_1-t_2)(\frac{1}{2m_F}\delta^{ij} 
\frac{\partial}{\partial t_1} -\epsilon^{ij})G_{++}(t_1-t_2,2m_F)\nonumber\\
&&\hskip -1in +n_F(m_F)(1-n_F(m_F))\left( {1\over (m_B+2m_F)^2}+ {1\over (m_B-2m_F)^2}\right)\cdot\nonumber\\
&&\hskip 2cm \cdot (\delta^{ij}-\frac{1}{m_B}\epsilon^{ij}
\frac{\partial}{\partial t_1})G_{++}(t_1-t_2,m_B)\nonumber\\
&&\hskip -1in +\left( {(1-n_F(m_F)-n_F(m_F+m_B))(1+n_B(m_B)-n_F(m_F))\over (m_B+2m_F)^2}\right.\nonumber\\
&& + \left. {(1-n_F(m_F)-n_F(m_F-m_B))(n_B(m_B)+n_F(m_F))\over (m_B-2m_F)^2}\right)\cdot\nonumber\\
&& \left.\cdot (\delta^{ij}+\frac{1}{2m_F}\epsilon^{ij}
\frac{\partial}{\partial t_1}) \left((\theta(t_1-t_2)-n_F(m_F))e^{-2im_F(t_1-t_2)} +(t_1\leftrightarrow  t_2)\right)\right]
\label{twoloop}
\end{eqnarray}

Note the appearance of terms proportional to $\delta^{ij}$ that are even under the interchange $t_1\leftrightarrow  t_2$, as well as the ``Chern-Simons'' terms proportional to $\epsilon^{ij}$ that are odd under the interchange $t_1\leftrightarrow  t_2$. It is also interesting to observe the new structure involving factors of $(t_1-t_2)$ multiplying propagators. The result (\ref{twoloop}) is non-zero even at zero temperature. This
should be compared and contrasted to the case in $2+1$ dimensional QED at zero temperature, where the parity-odd two-loop contribution to the gauge self-energy vanishes \cite{kao2,coleman}. Here this is not the case, a reflection of the fact that our dimensionally reduced model does not have local gauge invariance. 

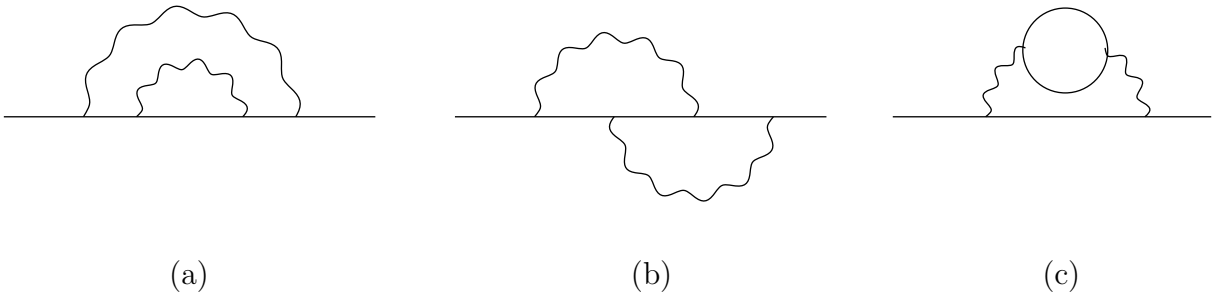
\begin{figure}
\centering{
\begin{picture}(460,140)(0,0)
\Line(5,80)(145,80)
\Line(175,80)(315,80)
\Line(340,80)(460,80)
\PhotonArc(75,80)(20,0,180)2 5
\PhotonArc(75,80)(40,0,180)2 6
\PhotonArc(235,80)(30,0,180)2 6
\PhotonArc(265,80)(30,180,360)2 6
\PhotonArc(405,80)(30,0,60)2 3
\PhotonArc(405,80)(30,120,180)2 3
\CArc(405,105)(16,0,360)
\Text(75,20)[c]{(a)}
\Text(250,20)[c]{(b)}
\Text(405,20)[c]{(c)}
\end{picture}}
\caption{The diagrams contributing to the two-loop fermion self-energy. The diagram in (b) vanishes due to gamma matrix properties (in fact, the 1-loop vertex correction itself vanishes - see caption to Figure 7).}
\label{f10}
\end{figure}

We have also computed the two-loop fermion self-energy, but the final expression is long and we do not present it here. Note that because of the vanishing of the one-loop vertex correction, the overlapping diagram in Figure \ref{f10}(b) is identically zero. The remaining two diagrams in Figure \ref{f10} are straightforward to evaluate given the one-loop results already presented. 

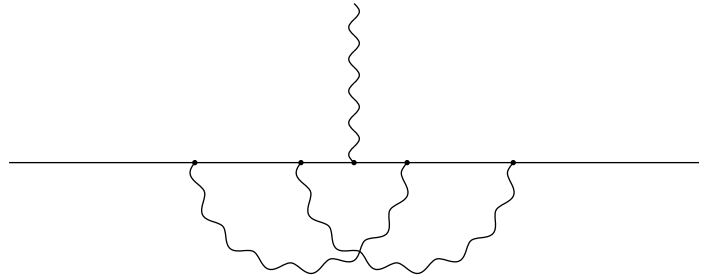
\begin{figure}
\centering{
\begin{picture}(300,130)(0,0)
\Line(20,70)(280,70)
\Photon(150,130)(150,70)2 5
\PhotonArc(170,70)(40,180,360)2 8
\PhotonArc(130,70)(40,180,360)2 8
\Vertex(130,70)1
\Vertex(210,70)1
\Vertex(90,70)1
\Vertex(170,70)1
\Vertex(150,70)1
\end{picture}}
\caption{The only non-vanishing diagram contributing to the two-loop vertex.}
\label{f11}
\end{figure}

Finally we mention that while the one-loop vertex correction (see Figure \ref{f7}) is zero, at the two-loop level there is in fact one non-zero vertex diagram, shown in Figure \ref{f11}. All other two-loop vertex diagrams vanish. 

\section{Conclusions} 
In this paper we have studied an extreme non-static limit of $2+1$-dimensional QED obtained by making a dimensional reduction so that all fields are spatially uniform (i.e. $\vec{p}=0$), but time dependent ($p^0\neq 0$). The resulting  $0+1$-dimensional field theory inherits many of the features of $QED_3$, such as Chern-Simons terms, time-reversal violation, an analogue of parity violation, and global $U(2)$ flavor symmetry. We found that at finite temperature the interactions induce a radiative Chern-Simons mass as well as a ${\cal T}$-even mass, and from the self-energy diagram we find that each is proportional to $\tanh(\frac{\beta m_F}{2})$, where $m_F$ is the fermion mass. Multi-leg diagrams contain non-extensive contributions at finite T, but it does not appear possible to re-sum exactly the effective action, except in the special case in which there are no bare ${\cal T}$-violating masses to begin with. This is indicative of the difficulty of analyzing the finite temperature $2+1$ model in non-static backgrounds. We have computed the two loop scalar and fermion self-energies, and we find that these are non-zero. Indeed, they are even non-zero at zero temperature. This does not violate the Coleman-Hill theorem \cite{coleman} since in our non-static limit we have suppressed local gauge invariance. It would be interesting to study supersymmetry breaking due to temperature effects in this model  - the model itself is not supersymmetric, but could be made so by the suitable inclusion of additional fields.

\vskip .5in
\noindent{\bf Acknowledgements}

This work was supported in part by US Department of Energy Grant No.
DE-FG-02-91ER40685 (AD) and Grant No. DE-FG-02-92ER40716 (GD). GD thanks V. Miransky for interesting discussions.

\begin{figure}
\centering{
\begin{picture}(200,200)(0,0)
\Photon(10,100)(70,100)4 4
\Photon(130,100)(190,100)4 4
\CArc(100,100)(30,0,180)
\CArc(100,100)(30,180,360)
\Vertex(70,100)1
\Vertex(130,100)1
\Text(0,100)[c]{i}
\Text(200,100)[c]{j}
\Text(75,100)[l]{a}
\Text(65,90)[c]{$t_1$}
\Text(135,90)[l]{$t_2$}
\Text(125,100)[r]{b}
\end{picture}}
\caption{The scalar self energy $\Pi_{ab}^{ij}(t_1-t_2)$ in coordinate space. The vertices are at coordinates $t_1$ and $t_2$, the external field indices are $i$ and $j$, and the thermal indices are $a$ and $b$.}
\label{f2}
\end{figure}
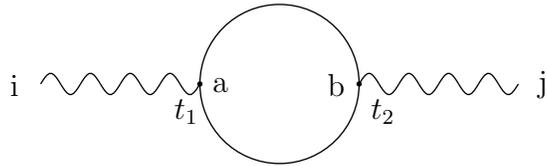

\begin{figure}
\centering{
\begin{picture}(400,120)(0,0)
\Text(10,50)[l]{$\Gamma_F[A_i]=$}
\Photon(70,50)(90,50)2 3
\Photon(110,50)(130,50)2 3
\LongArrowArc(100,50)(10,90,90)
\Text(150,50)[c]{+}
\Photon(170,50)(190,50)2 3
\Photon(210,50)(230,50)2 3
\Photon(200,60)(200,80)2 3
\Photon(200,40)(200,20)2 3
\LongArrowArc(200,50)(10,45,45)
\Text(270,50)[c]{$+\dots$}
\end{picture}}
\caption{Perturbative expansion of the effective action.}
\label{f3}
\end{figure}

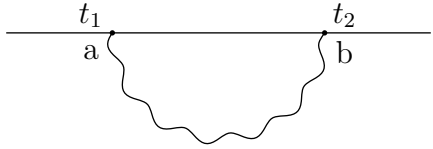
\begin{figure}
\centering{
\begin{picture}(200,100)(0,0)
\Line(180,80)(20,80)
\PhotonArc(100,80)(40,180,360)2 7
\Vertex(140,80)1
\Vertex(60,80)1
\Text(52,87)[c]{$t_1$}
\Text(148,87)[c]{$t_2$}
\Text(148,73)[c]{b}
\Text(52,73)[c]{a}
\end{picture}}
\caption{The fermion self energy $\Sigma_{ab}(t_1-t_2)$ in coordinate space. The vertices are at coordinates $t_1$ and $t_2$, and the thermal indices are $a$ and $b$.}
\label{f6}
\end{figure}

\begin{figure}
\centering{
\begin{picture}(400,200)(0,0)
\Photon(10,100)(70,100)4 4
\Photon(130,100)(190,100)4 4
\CArc(100,100)(30,0,180)
\CArc(100,100)(30,180,360)
\PhotonArc(100,142)(30,225,315)4 4
\Vertex(70,100)1
\Vertex(130,100)1
\Vertex(79,121)1
\Vertex(121,121)1
\Text(100,50)[c]{(a)}
\Photon(210,100)(270,100)4 4
\Photon(330,100)(390,100)4 4
\CArc(300,100)(30,0,180)
\CArc(300,100)(30,180,360)
\Photon(300,130)(300,70)4 4
\Vertex(270,100)1
\Vertex(330,100)1
\Vertex(300,130)1
\Vertex(300,70)1
\Text(300,50)[c]{(b)}
\end{picture}}
\caption{Two-loop contributions to the scalar self energy. The crossed diagram in (b) vanishes due to Dirac gamma matrix identities (in fact, the 1-loop vertex correction itself vanishes - see caption to Figure 7).}
\label{f8}
\end{figure}
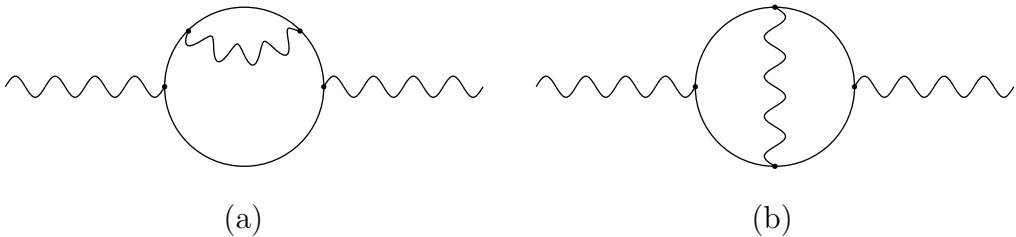

\end{document}